\documentclass[%
 reprint,
 superscriptaddress,
 amsmath,amssymb,
 aps,
 prx,
 nolongbibliography
]{revtex4-2}

\usepackage{graphicx}
\usepackage{dcolumn}
\usepackage{bm}

\usepackage{xcolor}

\begin{document}

\preprint{APS/123-QED}

\title{Motile bacteria crossing liquid-liquid interfaces}

\author{Jiyong Cheon}
\affiliation{Department of Physics, Ulsan National Institute of Science and Technology, Ulsan, Republic of Korea}

\author{Joowang Son}
\affiliation{Department of Physics, Ulsan National Institute of Science and Technology, Ulsan, Republic of Korea}

\author{Sungbin Lim}
\affiliation{Department of Biological Sciences, Ulsan National Institute of Science and Technology, Ulsan, Republic of Korea}

\author{Yundon Jeong}
\affiliation{Department of Biomedical Engineering, Ulsan National Institute of Science and Technology, Ulsan, Republic of Korea}

\author{Jung-Hoon Park}
\affiliation{Department of Biomedical Engineering, Ulsan National Institute of Science and Technology, Ulsan, Republic of Korea}

\author{Robert J. Mitchell}
\affiliation{Department of Biological Sciences, Ulsan National Institute of Science and Technology, Ulsan, Republic of Korea}

\author{Jaeup U. Kim}
\affiliation{Department of Physics, Ulsan National Institute of Science and Technology, Ulsan, Republic of Korea}

\author{Joonwoo Jeong}
\email{jjeong@unist.ac.kr}
\affiliation{Department of Physics, Ulsan National Institute of Science and Technology, Ulsan, Republic of Korea}

\date{\today}

\begin{abstract}
Real-life bacteria often swim in complex fluids, but our understanding of the interactions between bacteria and complex surroundings is still evolving. 
In this work, rod-like \textit{Bacillus subtilis} swims in a quasi-2D environment with aqueous liquid-liquid interfaces, \textit{i.e.}, the isotropic-nematic coexistence phase of an aqueous chromonic liquid crystal. 
Focusing on the bacteria motion near and at the liquid-liquid interfaces, we collect and quantify bacterial trajectories ranging across the isotropic to the nematic phase.
Despite its small magnitude, the interfacial tension of the order of 10 $\mathrm{\mu N/m}$ at the isotropic-nematic interface justifies our observations that bacteria swimming more perpendicular to the interface have a higher probability of crossing the interface. 
Our force-balance model, considering the interfacial tension, further predicts how the length and speed of the bacteria affect their crossing behaviors.  
We also find, as soon as the bacteria cross the interface and enter the nematic phase, they wiggle less, but faster, and that this occurs as the flagellar bundles aggregate within the nematic phase.
\end{abstract}

\maketitle
\section{Introduction}
Microbes are active colloids in complex environments \cite{Toner2005, Marchetti2013, Bechinger2016}. While consuming energy from their surroundings, these out-of-equilibrium objects move, grow, reproduce, and die. Their interactions with complex environments, ranging from flowing viscoelastic fluids and hard corrugated surfaces to anisotropic fluids with designed patterns \cite{Duchesne2015a, Mushenheim2014b, Zhou2014a, Mushenheim2014a, Chi2020, Sokolov2015, Peng2016, Turiv2020a}, not only impact their survival but also impact other living organisms and nature on Earth \cite{Chen2007, Valentine2010, Shunmugam2021, Wu2000, Josenhans2002, Ottemann2002, Zegado2023, Zhou2022}. These interactions also determine how microbes move and change in response to external stimuli and their spatiotemporal gradients \cite{ Zhou2022, Berg1972, Alon1998, Wadhams2004, Szurmant2004, Liebchen2018, Alert2022}. Understanding these phenomena benefits both our fight against, as well as our alliance with them.

Microbes interacting with interfaces, where the environment changes discontinuously, are of both fundamental and practical interest. Near or at interfaces, they exhibit different behaviors from the bulk ones. They can be scattered \cite{Makarchuk2019, Hoeger2021}, swim chirally via hydrodynamic interactions \cite{Lauga2006, Lauga2009, Lemelle2010, Lemelle2013, Lauga2016, Bianchi2019}, and form communities with emergent properties  \cite{Persat2015, Meylan2018}, such as resistance to physical disturbance and antibiotics. Microbes can even change the interfaces, and vice versa, \textit{e.g.}, biofilm formation on deformable soft substrates, and this mechanical feedback plays a critical role in their form and growth \cite{Asally2012, Zhang2017, Yan2018, JingYan2019}. However, unlike biofilm communities, a single microbe cannot deform typical fluid-fluid interfaces, including air-water or oil-water interfaces \cite{Lemelle2010, Bianchi2019, Deng2020}, but can be adsorbed onto them \cite{Vaccari2017, Deng2020}.

\begin{figure}[t!]
\centering 
\includegraphics[width=1\columnwidth]{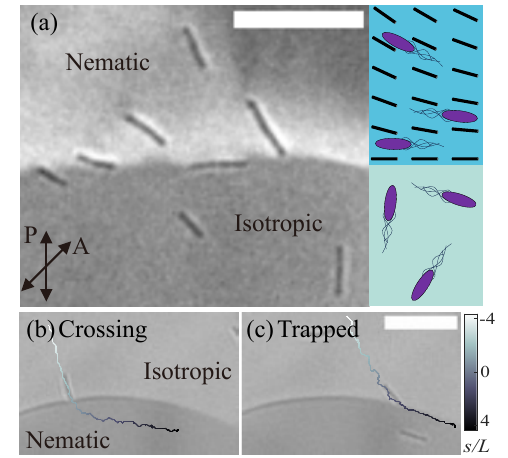}
\caption{
\textit{B. subtilis} in the nematic (N)-isotropic (I) coexistence of TB-DSCG. (a) Polarized optical microscopy (POM) image (left) of the bacteria in the coexistence phase alongside a schematic diagram (right) of the system. The analyzer (A) and polarizer (P) distinguish the birefringent N phase from the I phase; motile bacteria are dispersed in both phases. The black rods in the schematic diagram indicate the nematic directors which guide the swimming direction of the bacteria. (b and c) Representative trajectories of bacterial cells either crossing or being trapped at the interface. The color code represents the contour distances from the contact point to the interface, normalized by the body length (L) of the bacterium. The scale bars are 20 $\mathrm{\mu}$m.
}
\label{fig:microscopy}
\end{figure}

In this work, we report our experimental observation describing the interactions between a motile bacterium, \textit{Bacillus subtilis}, and an aqueous two-phase interface possessing a low interfacial tension of $\sim 10~\mathrm{\mu N/m}$. The bacterium in the quasi-two-dimensional isotropic-nematic coexistence phase of an aqueous lyotropic liquid crystal deforms, breaches, and crosses the interface while swimming from the isotropic to the nematic phase. Considering the force imparted on the bacterium by the deformed interface, we propose a minimal model to rationalize their crossing behavior's strong dependence on the incident angle to the interface, as reported in simulations \cite{Ryoichi2022} or in interface-free systems with a sharp viscosity gradient \cite{Coppola2021, Gong2022}. 
Additionally, we report immediate changes in the swimming behavior of the bacterium upon entering the nematic phase and propose that a nematic phase-induced change in the conformation of flagellar bundles may explain the observation \cite{Clopés2021}.

\begin{figure*}[t!]
\centering 
\includegraphics{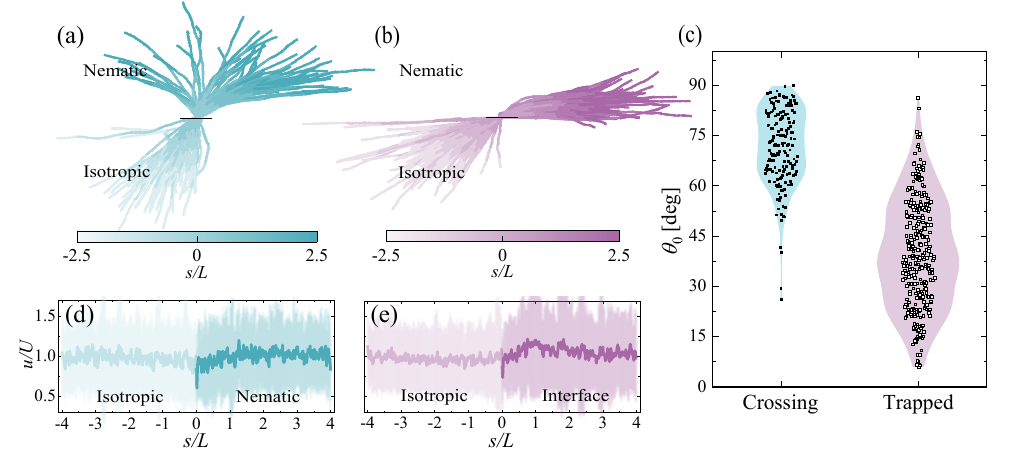}
\caption{
Interactions between the I-N interface and swimming bacteria.
(a) and (b) Overlapped trajectories of (a) 63 bacteria crossing the interface and (b) 119 bacteria being trapped at the interface. The trajectories are translated, rotated, and flipped to pass through the origin --- where the bacteria first meet the interface --- from the lower left; the black solid line at the origin represents the tangential line of the interface at the contact point. Note that the upper right curves in (b) reflect the curved interfaces. The color code represents the normalized travel distance as in Fig.~\ref{fig:microscopy}(b).
(c) Distribution of incident angles and their correlation to the crossing behavior. The scattered data points and the background band show the individual incident angles of 180 bacteria that crossed the interface and their density distribution, respectively; 297 trapped bacteria data is presented in the same way.
(d) and (e) Changes in the normalized speed of bacteria after contact with the interface. The solid lines in (d) and (e) are the average normalized speed profile for the crossed and trapped bacteria in (a) and (b). The instantaneous speed $u$ of each bacterium as a function of the normalized travel distance ($s/L$) from the contact point is divided by its average speed $U$ when far from the interface, \textit{i.e.}, $-4 < s/L < -3$.
}
\label{fig:trajectory}
\end{figure*}

\section{Results and Discussion}

As shown in Fig.~\ref{fig:microscopy}(a), polarized optical microscopy discerns the nematic and isotropic phase of the disodium cromoglycate (DSCG) dissolved in Terrific Broth (TB) and the rod-shaped \textit{B. subtilis} therein.
The schematic diagram sketches bacteria swimming in a quasi-2D liquid crystal environment with a wall-like interface present between the coexisting phases.
The bacteria-dispersed TB-DSCG is confined between two air-permeable substrates with the cell gap comparable to the diameter of a single bacterial cell, \textit{i.e.}, bacteria diameter $\sim$1.5 $\mathrm{\mu}$m and cell gap $<$ 5 $\mathrm{\mu}$m (see methods for more details). 
In fact, Fig.~\ref{fig:microscopy}(a) and Movie S1 show that all microbes in the field of view are in focus; they move freely in the $x-y$ plane but not along the $z$-direction.
Within the system, the bacteria in the nematic phase swim along the nematic directors that are parallel to the substrates and the isotropic-nematic interface. This boundary condition limits the crossing of a bacterium from the nematic to isotropic phase, except at topological defects \cite{Mushenheim2014a}.
However, in the isotropic phase, the bacteria swim isotropically and meet the liquid-liquid interface with any incident angles (see Fig.~S1).

\subsection{Interfacial tension matters in bacterial crossing behaviors from the isotropic to nematic phase}

The liquid-liquid interface affects the bacterial motion from the isotropic to the nematic phase, allowing a bacterium either to cross or to be trapped at the interface. 
Figure~\ref{fig:microscopy}(b), Movies S2 and S3 show representative cases, and Figs.~\ref{fig:trajectory}(a) and (b) overlap the trajectories of all investigated bacterial poles, \textit{i.e.}, the moving front.
While interacting with the interface, the rod-shaped bacteria may rotate. 
In such instances, if the angle between the bacterium and the interface's tangential line decreases gradually to zero before its travel distance from the contact with the interface reaches its body length, it remains at the isotropic phase and swims along the interface afterward (see Fig.~S2).
Otherwise, the bacterium will breach the interface and enter the nematic phase.
These observations hint that the liquid-liquid interface exerts a force on motile bacteria, affecting their eventual swimming trajectories.

In fact, bacteria with higher incident angles, \textit{i.e.}, more perpendicular to the interface, have a higher probability of crossing the interface. 
Figure~\ref{fig:trajectory}(c) summarizes the incident angle $\theta_0$ distributions, indicating bacteria that cross the interface and enter the nematic phase have $\theta_0 \gtrsim 60$ deg.
In contrast, when $\theta_0 \lesssim 60$ deg, the bacteria tended to become trapped at the interface.
This data aligns with our observation that the bacterial cell and their swimming direction rotate via an interaction with the interface and get trapped when they swim more parallel to the interface.
Notably, each data point comes from an individual bacteria cell, representing a range of body lengths and incident speeds. 
Intriguingly, we find no evidence of strong correlations between the crossing probability, body length, and incident speed (see Fig.~S3).

Changes in the bacterial swimming speeds at the interface corroborate the force encountered at the interface affects the bacteria's motion and their crossing behavior.
Figures~\ref{fig:trajectory}(d) and (e) show how the swimming speeds change while bacteria interact with the interface, according to whether they cross the interface. 
Because of the intrinsic heterogeneity observed in their motility, we normalize each bacterium's series data of instantaneous speed $u$ as a function of the contour distance $s$ along the trajectory with $L$ and $U$, respectively, where $L$ is the body length of the bacterium, and $U$ is its average `bulk' speed when far from the interface, \textit{i.e.}, $-4 < s/L < -3$.
The thick line corresponds to the average curve. 
As shown in Fig.~\ref{fig:trajectory}(d), the swimming speed slightly decreases as bacteria approach the interface, reaching a minimum when they encounter it. Subsequently, the speed recovers as they swim approximately one body length ($s/L \approx$  1) after breaching the interface.
A similar speed change after they meet the interface also occurs when bacteria get trapped. 
The origins of interactions between the bacteria and the interface deserve further investigation.

\begin{figure}
\centering
\includegraphics[width=1\columnwidth]{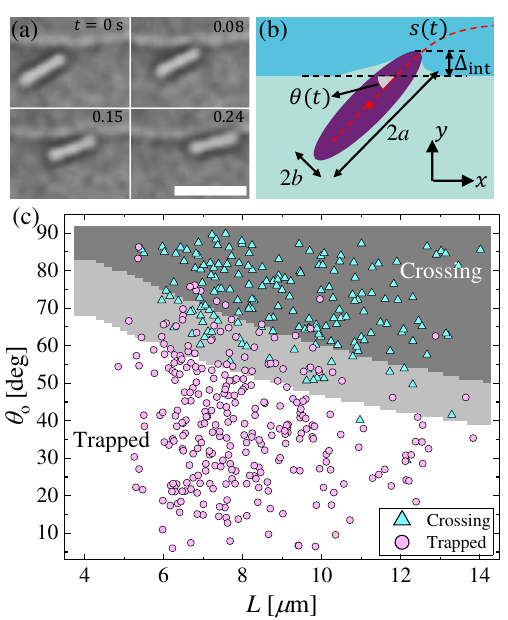}
\caption{
Our model and its comparison with the experimental data.
(a) A representative bacterium deforming the interface and getting trapped. (b) The schematic diagram of our model. The ellipsoid with major and minor axes, $2a$ and $2b$, represents a model bacterium deforming the interface by $\Delta_{\mathrm{int}}$, mimicking the bacterium at (a): $t$ = 0.08 and 0.15 s. Its center of mass (red dot) moves along the path $s$ in the $x-y$ plane while changing the angle $\theta(t)$ to the interface. The scale bar is 10 $\mu$m.
(c) Comparison between experimental data and model calculation. Each data point represents an observed bacterium with body length $L$ and incident angle $\theta_0$. The triangles designate \textit{B. subtilis} that cross the interface, which are shown in Fig.~\ref{fig:trajectory}(a), while the circles are bacteria that become trapped, in Fig.~\ref{fig:trajectory}(b). The model calculation results divide the background regions. The top gray region corresponds to the conditions for a model bacterium to cross the interface when its `bulk' speed is 30.3 $\mathrm{\mu}$m/s. The white region is for the ones that fail to cross the interface when the speed is 20.9 $\mathrm{\mu}$m/s.
}
\label{fig:model} 
\end{figure}

We propose that the force governing the observed crossing behavior in Fig.~\ref{fig:trajectory}(c) is the interfacial tension at the liquid-liquid interface.
The interfacial tension $\gamma$ at the isotropic-nematic interface is of the order of $10~\mathrm{\mu N}$/m \cite{Paparini_2021}, which is three orders of magnitude smaller than the interfacial tension of a typical oil-water interface. 
Even though the tension of this aqueous two-phase system is small, with interfacial deformation at a micron scale, it can exert a force, \textit{i.e.}, $10~\mathrm{\mu N/m} \times 1~\mathrm{\mu m} \sim 10~\mathrm{pN} $, that is comparable to the \textit{B. subtilis}'s propulsion force $\sim 10~\mathrm{pN}$ (see Fig.~S3(c)).
Indeed, Fig.~\ref{fig:model}(a) shows the bacterium's moving front protrudes from the isotropic phase, displacing the interface by approximately $1~\mathrm{\mu m}$ before it gets trapped at the interface. 
This observation hints that the interface resists deformation by motile bacteria, pushes back on the bacteria like a spring, and affects their trajectories and crossing of the interface.
We believe that the liquid crystallinity of the nematic phase resisting deformation may contribute in a similar way, possibly to a similar or smaller extent. 
This may be because low-concentration and high-temperature DSCG, \textit{i.e.}, our DSCG at the coexistence phase, possesses small elastic moduli $\sim$ 1 pN \cite{Zhou014}.

Our theoretical model provides the equations of motion to describe the bacteria interacting with the interface. 
A prolate ellipsoid, as sketched in Fig.~\ref{fig:model}(b), simplifying a bacterium, swims along its major axis with a constant propulsion force and no stochastic tumbling. 
Considering our quasi-2D confinement, it translates and rotates only in a 2D plane.
We then represent the interface as a straight line parallel to the $x$-axis, which divides the isotropic and nematic phases. 
When a bacterium in the isotropic phase encounters the interface, as shown in Fig.~\ref{fig:model}(b), the deformed interface exerts a normal force, \textit{i.e.}, parallel to the $y$-axis, on the bacterium.
Adopting Hooke's law, we approximate that the force is proportional to the interfacial tension $\gamma$ multiplied by the deformed height $\Delta_{\mathrm{int}}$ along $y$.
We also ignore the effects from the top and bottom substrates and approximately adopt the translational and rotational drag coefficients of a prolate ellipsoid in 3D, $\mu_{\mathrm{trans}} = \frac{4\pi\eta a}{\ln(2a/b)-1/2}$ and $\mu_{\mathrm{rot}} = \frac{8\pi\eta a^3 /3}{\ln(2a/b)-1/2}$ \cite{Berg1984}, respectively, with semi-major axis $a$, semi-minor axis $b$, and bulk viscosity $\mu$. Note that we assume the viscosities in the isotropic and nematic phases are very similar since the average swimming speeds barely change after crossing the interface, as shown in Fig.~\ref{fig:trajectory}(d). Furthermore, based on the near perpendicular contact angle of the interface on the bacterial body, we exclude the possibility that the bacteria have an affinity to a certain phase (see Fig.~S4). All these observations, including the small interfacial tension $\gamma$, imply these two phases are very similar.

We write and solve the following equations of bacteria's overdamped motion.
\begin{align}
\mu_{\mathrm{trans}}~\frac{d y_{\mathrm{B}}(t)}{d t} = F_{\mathrm{prop}} \sin \theta(t) - c \gamma \Delta_{\mathrm{int}} \sin^2 \theta(t)\label{eq:translation}\\
\mu_{\mathrm{rot}} \frac{d \theta(t)}{d t} = -a c \gamma \Delta_{\mathrm{int}} \cos \theta(t).
\label{eq:rotation}
\end{align}
In Fig.~\ref{fig:model}(b), the red dot represents the position of the ellipsoid's center of mass ($x_{\mathrm{B}}(t)$,~$y_{\mathrm{B}}(t)$), and $\theta(t)$ is the angle between the ellipsoid and interface.
$F_{\mathrm{prop}}$ is the bacteria's propulsion force, and we use $F_{\mathrm{prop}} = \mu_{\mathrm{trans}} V$, where $V$ is the ellipsoid's speed, which is assumed to be constant.
The force from the deformed interface is $c \gamma \Delta_{\mathrm{int}}$ with a proportional coefficient $c$, the interfacial tension $\gamma$, and the interfacial deformation $\Delta_{\mathrm{int}} = y_{\mathrm{B}}(t) + a \sin(\theta(t))$ when the undeformed interface is at $y = 0$.
We write another equation for the rotational motion considering the torque by the interfacial force, Eq.~(\ref{eq:rotation}). See SI for the derivation.

Solving the set of differential equations, Eqs.~(\ref{eq:translation}) and (\ref{eq:rotation}), we find the maximum $\Delta_{\mathrm{int}}$ to determine whether an ellipsoid will cross the interface.
We choose model parameters based on experimental observations and literature values and use $\gamma$ = 10 $\mathrm{\mu}$N/m, $\eta$ = 30 mPa$\cdot$s \cite{Duchesne2015a}, $b$ = 0.7 $\mathrm{\mu}$m, $a$ = 2 -- 7 $\mathrm{\mu}$m, and $V$ = 15 -- 35 $\mathrm{\mu}$m/s.
We adopt $c$ = 1 considering that $\gamma$ = 10 $\mathrm{\mu}$N/m is already an order-of-the-magnitude estimation.
For the initial conditions, \textit{i.e.}, when the ellipsoid's tip meets the interface, we employ $y_{\mathrm{B}}(t = 0) = -a \sin(\theta(t = 0))$ and $\theta(t = 0) = \theta_{0}$ = 0 -- 90 deg.
We hypothesize that the ellipsoid can cross the interface only when $\Delta_{\mathrm{int}}$ surpasses a critical value.
As shown in Fig.~\ref{fig:model}(a), our experiment indicates that the critical $\Delta_{\mathrm{int}}$ for a bacterium to cross the interface is of the order of 1~$\mathrm{\mu}$m.
Namely, for each ($a$, $V$, $\theta_0$), we numerically calculate the maximum $\Delta_{\mathrm{int}}$ along the ellipsoid's trajectory and check if it is greater than the critical value, \textit{i.e.}, 1~$\mathrm{\mu}$m.

As shown in Fig.~\ref{fig:model}(c), our model is supported by the experimentally observed crossing behavior, such as displaying strong angle dependency.
The scattered experimental data overlap well with the model-predicted colored regions.
The triangles represent bacteria that cross the interfaces and lie mostly in the dark gray region, where the model predicts that the interfacial deformation surpasses the critical value when the swimming speed of the model ellipsoid $V$ = $U_{\mathrm{avg}} + U_{\mathrm{stdev}}$. Here, $U_{\mathrm{avg}}$ = 25.6~$\mathrm{\mu}$m/s, and  $U_{\mathrm{stdev}}$ = 4.7~$\mathrm{\mu}$m/s represent the ensemble average and the standard deviation of bacterial speed $U$ from the experimentally measured trajectories (see Fig.~S3). 
The boundary between the light gray and white regions is calculated using $V$ = $U_{\mathrm{avg}} - U_{\mathrm{stdev}}$, manifesting the speed-dependency of the crossing behavior. Note each data point is from a different bacterium's trajectory with various $U$, and see Fig.~S5 for a $\theta_0-U$ plot.

Crossing behavior's weak dependency on the body length and speed in the model hints at why we observe a strong correlation only between the incident angle and the crossing probability, as shown in Fig.~\ref{fig:trajectory} and Fig.~S3.
It seems that the distributions of body length $L = 8.0~\pm~2.0~\mathrm{\mu}$m and swimming speed $U = 25.6~\pm~4.7~\mathrm{\mu}$m/s are not broad enough to result in experimentally observable dependencies. Furthermore, they likely contribute to the crossing behavior in complicated ways, via the propulsion force and drag coefficient.
In contrast, as shown in Fig.~S1, the isotropic environment allows the bacteria to explore all possible incident angles to the interface. 

\subsection{Bacteria wobble less but faster in the nematic phase}

\begin{figure*}
\centering
\includegraphics{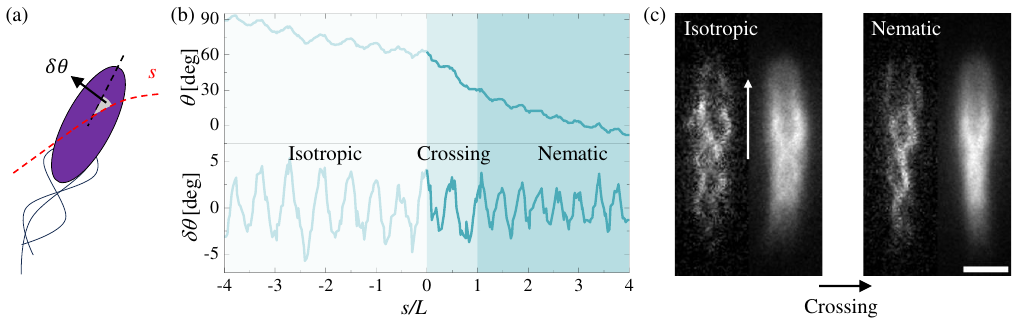}
\caption{
Wiggling trajectories and flagella of bacteria crossing the isotropic-nematic interface.
(a) Schematic diagram of bacterial wiggling in a quasi-2D environment. The angle between the axis of the bacterial body and the tangential line defining the bacterial center's path $s$ is $\delta \theta$. 
(b) A representative trajectory shown with the angle $\theta$ between the body and interface (top) and $\delta \theta$ (bottom) as a function of the normalized travel distance $s/L$. The shaded backgrounds are included to distinguish the phases.
(c) Comparison of the fluorescently labeled flagella of a representative bacterium in the isotropic and nematic phase. The left and right images in each phase correspond to a snapshot and a 20-frame-averaged image of a spinning bacterium's fluorescently labeled flagella, respectively. See Movies S5 and S6. These images are acquired from the same bacterium before and after it crosses the interface. The white arrow indicates the swimming direction. The scale bar is 2 $\mathrm{\mu m}$.
}
\label{fig:wiggle&bundle} 
\end{figure*}

Comparing bacterial motions in the isotropic and nematic phases, we find that bacterial oscillations, i.e., wiggles, are less intense (reduced amplitude) but more frequent in the nematic phase. 
Fig.~\ref{fig:wiggle&bundle}(b) shows representative data of how the angle $\theta$ between a bacterium and the interface changes as the bacterium crosses the interface. The high-frequency oscillations in the slowly varying $\theta$ profile result from wiggling, which is depicted here as a 2D projection of a bacterium's 3D wobbling motion, as sketched in Fig.~\ref{fig:wiggle&bundle}(a).
After extracting the oscillation part shown in the bottom graph of Fig.~\ref{fig:wiggle&bundle}(b), the average oscillation amplitude and wavelength are measured and used to define the wiggling angle $\phi$ and pitch, respectively, in both the isotropic and nematic phases.
Similarly, we measure the wiggling frequency in the time domain (see Fig.~S6).
As shown in Fig.~\ref{fig:wiggle&bundle}(b), the wiggling angle and pitch, \textit{i.e.}, the oscillation amplitude and wavelength, decrease as soon as the bacterium enters the nematic phase, while the frequency increases.
The bacterium still swims at a similar speed as in the isotropic phase when swimming within the nematic phase, as shown in Fig.~\ref{fig:trajectory}(d), because the decrease in wiggling pitch is compensated by the increase in frequency.
The investigation of 56 wiggling trajectories tells that the wiggling angle, pitch, and frequency in the isotropic phases are $3.7 \pm 1.7$ deg, $5.5 \pm 1.1~\mu$m, and $4.6 \pm 1.5$ Hz, respectively. Then, entering the nematic phase changes the wiggling angle by $-0.8 \pm 1.0$ deg, the pitch by $-1.2 \pm 1.2~\mathrm{\mu m}$, and the frequency by $1.9 \pm 1.7$ Hz (see Fig.~S7).

To investigate why the oscillatory motion changes after crossing the interface, the bacterial flagella are visualized by fluorescently labeling them. As shown in Fig. \ref{fig:wiggle&bundle}(c) and Movie S4, we find the flagellar bundles aggregate as a bacterium enters the nematic phase. We use a smooth-swimming mutant (mutated for \textit{cheB} gene; DK2178) for the flagella visualization \cite{Kris2008}. 
The peritrichous flagella, which are uniformly distributed over the bacterial body, make bundles and generate propulsion in the isotropic phase, as shown in Fig.~\ref{fig:wiggle&bundle}(c, left) and Movie S5.
However, as the bacteria enter the nematic phase, the dispersed bundles aggregate and rotate together, as shown in Fig.~\ref{fig:wiggle&bundle}(c, right) and Movie S6.
Comparing the time-averaged images in the isotropic and nematic phases corroborates this observation: the flagella in the nematic phase are seen to align together into a narrower aggregate, particularly at their tail ends.

We propose that elasticity-mediated attractive interactions between flagellar bundles in the nematic phase result in their aggregation, and that this leads to the observed shifts in the oscillatory behavior.
The flagella in the nematic phase tend to align parallel to the nematic directors~\cite{Goral2022}.
Presumably, this alignment is energetically favorable as it disturbs the nematic alignment less.
Furthermore, we assume the aggregation of the bundles similarly minimizes the energetic penalty, giving rise to an effective attractive interaction between bundles.
We also find the reduction in bundle numbers alleviates the misalignment between the axis of flagellar rotation and that of the bacterial body, resulting in less wiggling~\cite{Clopés2021}.
Lastly, we note that the wiggling frequency increases and the pitch decreases after entering the nematic phase, leading to almost no change in the swimming speed, a finding that deserves further investigation.

\section{Conclusion}
In summary, our \textit{B. subtilis} dispersed in the isotropic-nematic coexistence phase can cross the liquid-liquid interface where the interfacial tension is $\sim 10~\mathrm{\mu N/m}$, mainly depending on their incident angle to the interface. We propose a model considering the interfacial deformation by bacteria's propulsion force. The model supports the strong incident-angle dependency and explains why experiments show no strong correlation between crossing behaviors and other motility parameters, such as incident speeds and bacterial body lengths. 
We also suggest that the nematic phase induces the aggregation of flagellar bundles and that these act to lessen the wiggling angle and pitch of \textit{B. subtilis}, while increasing its frequency.

Our findings and future work may shed new light on transport phenomena in complex environments. First, we expect tiny interfacial tensions present in liquid-liquid phase separation may impact transport phenomena associated with active matter, not only at the microbial scales but also at macromolecular ones, as our results show this depends on interfacial deformation and active force. Furthermore, beyond the current model's simple interfacial deformation, the roles of affinity asymmetry and complex fluid properties deserve further investigation. For instance, if a microbe strongly favors one phase over the other, additional energetics should be considered in their transport and partitioning. In a similar vein, it seems probable that fluids' complex properties, \textit{e.g.}, the liquid crystallinity or non-Newtonian properties, affect the transport phenomena. 
In fact, the elasticity of the nematic phase seems to affect the conformation of flagellar bundles, resulting in the altered wiggling motion of bacteria observed.

\section*{Acknowledgment}
Authors thank Daniel B. Kearns for providing \textit{B. subtilis} DK2178. We acknowledge the financial support from the National Research Foundation of Korea through NRF-2020R1A4A1019140. J.C. and J.J. acknowledge NRF-2021R1A2C101116312. J.S. and J.U.K. acknowledge RS-2023-00257666. S.L. and R.J.M. acknowledge NRF-2020R1A2C2012158. Y.J. and J.P. acknowledge NRF-2021R1A2C3012903.

\bibliography{References}

\end{document}



\title{Supplemental Material: Motile bacteria crossing liquid-liquid interfaces}

\author{Jiyong Cheon}
\affiliation{Department of Physics, Ulsan National Institute of Science and Technology, Ulsan, Republic of Korea}

\author{Joowang Son}
\affiliation{Department of Physics, Ulsan National Institute of Science and Technology, Ulsan, Republic of Korea}

\author{Sungbin Lim}
\affiliation{Department of Biological Sciences, Ulsan National Institute of Science and Technology, Ulsan, Republic of Korea}

\author{Yundon Jeong}
\affiliation{Department of Biomedical Engineering, Ulsan National Institute of Science and Technology, Ulsan, Republic of Korea}

\author{Jung-Hoon Park}
\affiliation{Department of Biomedical Engineering, Ulsan National Institute of Science and Technology, Ulsan, Republic of Korea}

\author{Robert J. Mitchell}
\affiliation{Department of Biological Sciences, Ulsan National Institute of Science and Technology, Ulsan, Republic of Korea}

\author{Jaeup U. Kim}
\affiliation{Department of Physics, Ulsan National Institute of Science and Technology, Ulsan, Republic of Korea}

\author{Joonwoo Jeong}
\email{jjeong@unist.ac.kr}
\affiliation{Department of Physics, Ulsan National Institute of Science and Technology, Ulsan, Republic of Korea}

\date{\today}

\keywords{liquid-liquid interface, microbe, cross}
\maketitle

\renewcommand\thefigure{S\arabic{figure}}
\renewcommand{\theequation}{S\arabic{equation}}

\section{Description of Supplementary Movies}

\textbf{Movie S1:} Motile \textit{Bacillus subtilis} dispersed in the isotropic-nematic coexistence phase with multiple interfaces. All bacteria in this large field of view are in focus, suggesting that this is a quasi-2D system. Scale bar = 50\,$\mu$m
\singlespacing
\textbf{Movie S2:} \textit{Bacillus subtilis} is crossing the liquid-liquid interface. Scale bar = 20~$\mu$m. The trajectory shows the positions of the moving front.
\singlespacing
\textbf{Movie S3:} \textit{Bacillus subtilis} is trapped at the liquid-liquid interface. Scale bar = 20~$\mu$m. The trajectory shows the positions of the moving front.
\singlespacing
\textbf{Movie S4:} \textit{Bacillus subtilis} with fluorescently-labeled flagella crossing the liquid-liquid interface. Scale bar = 5~$\mu$m.
\singlespacing
\textbf{Movie S5:} \textit{Bacillus subtilis} with fluorescently-labeled flagella swimming in the isotropic phase is shown in a body-centered frame of reference. The arrow indicates the swimming direction. Scale bar = 2~$\mu$m.
\singlespacing
\textbf{Movie S6:} \textit{Bacillus subtilis} with fluorescently-labeled flagella swimming in the nematic phase is shown in a body-centered frame of reference. The arrow indicates the swimming direction. Scale bar = 2~$\mu$m.
\singlespacing

\newpage
\section{Materials and Methods}

\subsection{Bacterial cell culture} 
A colony of \textit{Bacillus subtilis} strain ATCC 6051 was first cultured overnight on a lysogeny broth (LB) agar plate at $37^{\circ}$C, transferred into the TB (Terrific Broth), and cultured in a shaking chamber set at $37^{\circ}$C and 250 rpm. After growing, the bacterial culture was diluted in fresh TB to an optical density (OD; 600 nm) of approximately 0.03 and cultured under the same conditions until the OD600 was 0.3.
Then, a bacterial suspension of OD600 = 1 in fresh TB was prepared by centrifuging the culture under 8k RCF for 1 minute at room temperature and resuspending the pellet into fresh TB. We then mixed this bacterial suspension with TB-DSCG right before the experiment.

\subsection{Flagella labeling}
\textit{Bacillus subtilis} strain DK2178 from Daniel B. Kearns' group \cite{Kris2008} was cultured in the same way as described in the literature. To minimize unnecessary chemical reactions during fluorescent labeling, we replaced the culture media (TB) of 1~ml bacterial suspension with 200~$\mathrm{\mu l}$ of phosphate-buffered saline (PBS). Then, 5~$\mathrm{\mu l}$ of dye solution (5~mg/ml of Alexa Fluor 488 C5 maleimide in DMSO)\cite{Kris2008} was added. After incubation in a shaking chamber for 20~min ($37^{\circ}$C and 250~rpm), most of unreacted dye was removed by pelleting the bacteria via centrifugation (11.4k RCF for 1 minute). Note that we conducted this centrifugation step only once because the difference in the background fluorescence signal from the remaining dye helped distinguish the nematic and isotropic phases. After washing, the bacteria were resuspended in fresh TB to a final OD600 of 1. We then mixed this bacterial suspension with TB-DSCG right before the experiment.

\subsection{Sample preparation}
Disodium cromoglycate (DSCG, $95\%$ purity) was purchased from Sigma-Aldrich and used with no further purification. 
A 16\% (wt/wt) TB-DSCG solution was prepared by dissolving the DSCG powder into autoclaved, sterile TB and incubating the solution at $65^{\circ}$C until complete dissolution of the DSCG. After cooling to room temperature, a bacterial suspension of OD600 = 1 in fresh TB was added to make a 12\% (wt/wt) TB-DSCG solution at the nematic phase. 
Approximately 1.3~$\mathrm{\mu l}$ of the bacterial TB-DSCG solution was sandwiched between two polymeric air-permeable coverslips (Cat. No. 20009, SPL Life Science) with no spacers, to create a quasi-2D environment with a thickness of less than 5~$\mathrm{\mu m}$. 
We approximate the cell gap from the known volume and spreading area of the solution, assuming a uniform thickness. We then sealed the edges of the sample using epoxy glue to minimize evaporation during the observation.

\subsection{Optical microscopy} 
We observed the bacterial body using an upright microscope (BX53-P; Olympus) equipped with a temperature-controlled stage (T95-PE120; Linkam Scientific Instruments) set to 30$^{\circ}$C, where the 12 \% (wt/wt) TB-DSCG solution is at the nematic and isotropic coexistence phase. We used a 40x dry objective and a CCD camera (INFINITY3-6URC; Lumenera) to capture the motion of bacteria at 70 frames per second.

Fluorescence microscopic images were acquired using a wide-field imaging setup composed of a microscope (IX73, Evident), an oil immersion objective lens (UPLAPO100XOHR, Evident), a dichroic mirror (ZT488rdc, Chroma), an emission filter (ET525/50m, Chroma), and a scientific CMOS camera (C14440-20UP, Hamamatsu) attached to the camera port of the microscope. A diode laser (Cobolt 06-MLD, 488 nm, 200 mW, Cobolt) was used as the excitation light source. We adopted a live streaming mode for the image acquisition, with a frame rate of 46.5 Hz and an exposure time of 20 ms.

\subsection{Deep learning-based tracking of motile bacteria} 
In order to extract the trajectory and body orientation of each bacterium from the recorded video data, we developed a deep learning-based object-tracking scheme. The first step of this process is embedding-based instance segmentation using the EmbedSeg network \cite{LALIT2022102523}. This proposal-free segmentation not only detects multiple \textit{B. subtilis} cells in each video frame but also enables us to maintain the ID of each bacterium even when it is close to another bacterium or overlaps with the liquid-liquid interface. As a result of the segmentation, pixel sets are assigned so that each set indicates one of the multiple bacterium instances in the frame. The position of a cell at time $t$ is then defined as the center of mass of the corresponding pixel set, and its body orientation is the direction of the smallest circumscribed rectangle of the pixels.

\subsection{Data availability} 
We provide the data that supports the findings of this study.
The body lengths and the incident angles of 477 bacteria are measured and specified by ID: C001 -- C180 (crossing) and T001 -- T297 (trapped). They are used in Figs.~2(c), 3(b), \ref{fig:incident angle distribution}, \ref{fig:angle change}(c-d), and \ref{fig:scatter plots}(a).
The trajectories of 303 bacteria are also presented in a similar way: C001\_txyth -- C110\_txyth (crossing trajectories) and T001\_txyth -- T193\_txyth (trapped trajectories). Among the 303 trajectories, the 182 trajectories that are long enough to have $|s/L| \geq 4$ are used in Figs.~2(a-b, d-e), \ref{fig:angle change}(a-b), \ref{fig:scatter plots}(b-d), and \ref{fig:angle vs speed}: C001\_txyth -- C063\_txyth and T001\_txyth -- T119\_txyth. The data in Fig.~4(a) and \ref{fig:Wiggling time domain} correspond to the C002\_txyth, and the data in Figs.~\ref{fig:Wiggling change}(a-c) are from C001\_txyth -- C056\_txyth.

\section{Supplementary Data} 

\subsection{Analysis of angles between bacteria and interfaces} 

\begin{figure}[htb!]
\centering 
\includegraphics{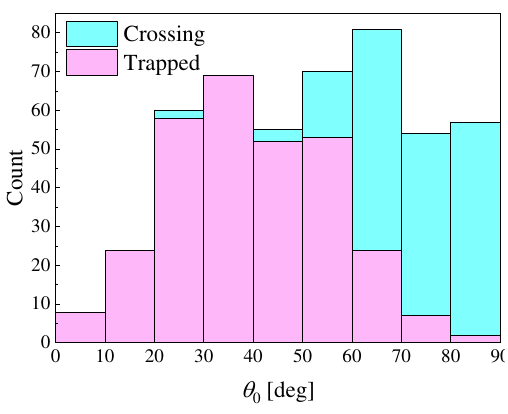}
\caption{Stack histogram of all incident angles of \textit{Bacillus subtilis} encountering the isotropic-nematic interface from the isotropic phase. We present the same data set used in Fig.~2(c): $N = 180$ (crossing) and $N=297$ (trapped).
}
\label{fig:incident angle distribution}
\end{figure}

\begin{figure}
\centering 
\includegraphics{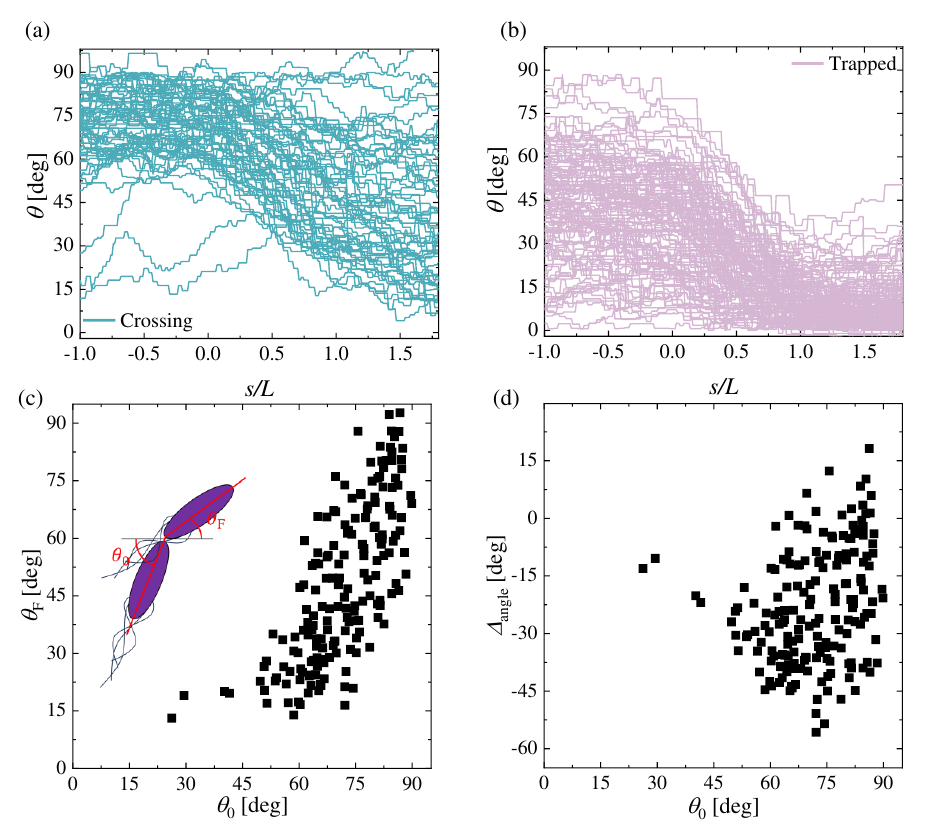}
\caption{Rotation of the bacteria interacting with the interface. (a) and (b) Change of the angle between bacteria and interfaces as a function of normalized travel distance. An individual trajectory is from a bacterium (a) crossing the interface and (b) being trapped at the interface. We present the same data set used in Figs.~2(a-b, d-e): $N = 63$~(crossing) and $N = 119$~(trapped)). (c) The angle before and after crossing interfaces ($N = 180$, the data set used in Fig.~2(c)). The incident angle is the angle when a bacterium meets the interface, and the final angle is estimated when the bacterium swims its body length after breaching the interface, as sketched in the inset. (d) The change in the angle, $\mathit{\Delta}_{\mathrm{angle}}=$ $\theta_{\mathrm{F}}-\theta_{0}$, as a function of the incident angle.}
\label{fig:angle change}
\end{figure}

We investigated 477 bacteria encountering the interface from the isotropic phase and found no preferred angle but a rather uniform distribution, as shown in Fig.~\ref{fig:incident angle distribution}. Namely, except for the very shallow angles, the total counts for various incident angles are similar, while the crossing behavior strongly depends on the incident angle. Figure~\ref{fig:angle change} shows how bacteria rotate as they move and interact with the isotropic-nematic interfaces. The angles of bacteria that successfully cross the interface typically decrease, as shown in Figs.~\ref{fig:angle change}(a, c-d). The angles of the trapped bacteria also decrease and converge to 0, \textit{i.e.}, parallel to the interface.

\newpage
\subsection{Crossing behavior's dependencies on other motility parameters}

\begin{figure}[htb!]
    \centering
    \includegraphics{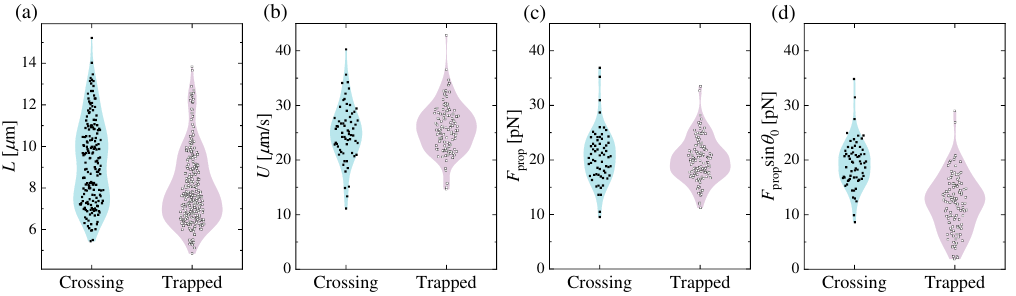}
    \caption{Crossing behavior's dependencies on other motility parameters: (a) body length, (b) incident speed, (c) propulsion force, and (d) propulsion force normal to the interface. The data in (a) is from the data set used in Fig.~2(c): $N = 180$ (crossing) and $N = 297$ (trapped), while the data in (b), (c), and (d) are from the data set used in Figs.~2(a-b): $N = 63$ (crossing) and $N=119$ (trapped).}
    \label{fig:scatter plots} 
\end{figure}

As shown in Fig.~\ref{fig:scatter plots}, we find no apparent correlations between crossing behavior and other motility parameters but the incident angle (see Fig.~2(c)) and the normal propulsion force, in which the incident angle contributes.

\subsection{The near 90-deg contact angle between the isotropic-nematic interface and \textit{B. subtilis}}

\begin{figure}[htb!]
    \centering
    \includegraphics{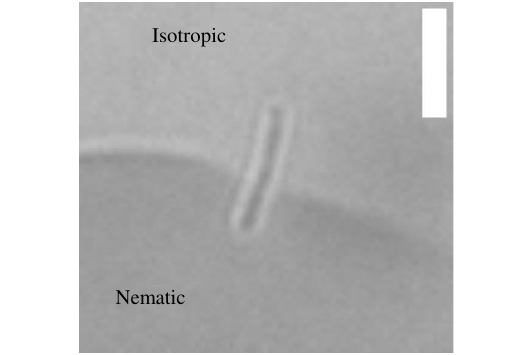}
    \caption{The 90-deg contact angle between the isotropic (I)-nematic (N) interface and immotile \textit{B. subtilis}. Scale bar = 10~$\mu$m}
    \label{fig:90deg contact angle} 
\end{figure}

There exist immotile \textit{B. subtilis} in the sample, possibly because they were stuck to the substrates. We find several immotile bacteria lying across the isotropic-nematic interface and discover that they all have a near 90-degree contact angle to the interface, as shown in Fig.~\ref{fig:90deg contact angle}. This indicates that \textit{B. subtilis} do not have an affinity to a specific phase, \textit{i.e.}, like each phase equally. This enables us to exclude affinity-related forces and assume that the interfacial tension and critical deformation of the interface determine the crossing behavior.

\subsection{Derivation of equations of motion} 

The ellipsoid in Fig.~3(b) is our model bacterium, which swims in the $x-y$ plane. We use the contour-length coordinate $s$ along which the center of the ellipsoid moves and assume the major axis of the ellipsoid is parallel to the tangential direction of the swimming path.

When the ellipsoid touches the interface parallel to the $x$ direction at time $t = 0$, the interface starts to deform and exert a force on the tip of the ellipsoid along the $y$ direction, which is normal to the interface. We assume this force $F_{\mathrm{{interface}}}$ is $c \gamma \Delta_{\mathrm{int}}$ with the interfacial tension $\gamma$, interfacial deformation $\Delta_{\mathrm{int}}$, and a proportional coefficient $c$, which is set to 1 considering that we use the order-of-magnitude estimation of $\gamma$. This force competes with the propulsion force $F_{\mathrm{prop}}$ and induces the rotation of the ellipsoid.

The overdamped equations of motion regarding the translational motion of the ellipsoid is 
\begin{equation} \label{trans}
0 = F_{\mathrm{prop}} - \mu_{\mathrm{trans}} \frac{ds_{\mathrm{B}}(t)}{dt} - F_{\mathrm{interface}} \sin\theta (t),
\end{equation}
where the translation drag coefficient $\mu_{\mathrm{trans}}$ of the ellipsoid along the major axis is $\frac{4\pi\eta a}{\ln(\frac{2a}{b})-\frac{1}{2}}$ with the semi-major and semi-minor axes, $a$ and $b$, respectively. The rotational motion is similarly governed by
\begin{equation} \label{rot}
0 = a F_{\mathrm{interface}} \cos\theta(t) - \mu_{\mathrm{rot}} \frac{d\theta(t)}{dt},
\end{equation}
where the rotational drag coefficient $\mu_{\mathrm{rot}}$ of the ellipsoid about the minor axis is $\frac{8\pi\eta a^3 /3}{\ln(\frac{2a}{b})-\frac{1}{2}}$.
Utilizing $ds \sin \theta = dy$, we multiply $\sin\theta$ on both sides of Eqs.~(\ref{trans}) and (\ref{rot}) and derive Eqs.~(1) and (2) of the main text.

\subsection{Comparison between experimental data and model calculation: Incident angle and speed} 

The experimental data set in Fig.~3(c) is shown again with model predictions according to the incident angle $\theta_0$ and speed $U$. Most triangles are located in the upper gray regions with higher incident angles, whereas the limited $U$-range and various body lengths $L$ give no clear evidence of $U$-dependencies.

\begin{figure}[htb!]
    \centering
    \includegraphics{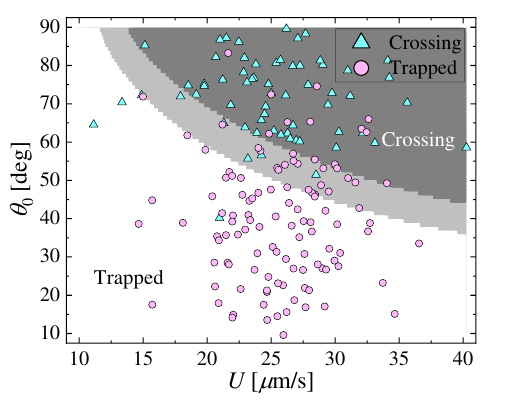}
    \caption{The comparison between experimental data and model calculation according to the incident angle and speed. Each data point is from the observation of an individual bacterium. The model calculation results divide the background regions. The top gray region corresponds to the condition for a model bacterium to cross the interface when its body length is longer than 10.0 $\mu$m. The bottom white region is when the bacteria fail to cross the interface when its body length is shorter than 6.1 $\mu$m. Here we present the data set used in Figs.~2(a) and (b): $N = 63$ (crossing) and $N=119$ (trapped), respectively.}
    \label{fig:angle vs speed} 
\end{figure}

\subsection{Analysis of wiggling trajectories}
\begin{figure}[htb!]
    \centering
    \includegraphics{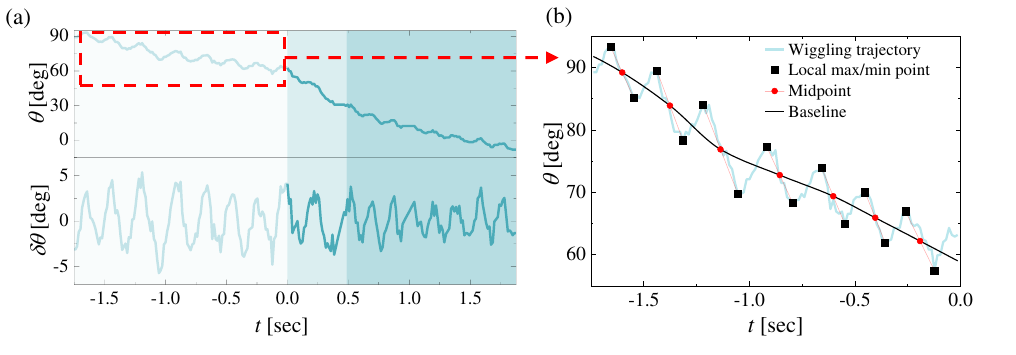}
    \caption{Analysis of a wiggling trajectory in the time domain. 
    (a) A representative wiggling trajectory before and after baseline subtraction. The top graph shows the angle $\theta$ between the body and interface as a function of time $t$, while the bottom plots $\delta \theta$ after the baseline subtraction. The same trajectory in the spatial domain is shown in Fig. 4(b). The shaded backgrounds are provided to help distinguish the phases in which a bacterium is located. 
    (b) Estimation of the baseline. The data within the red box of (a) is magnified here. The black squares indicate the local minima/maxima of $\theta(t)$, and the red circles between the connected squares are their midpoints. Interpolating the red circles into a solid black line gives the baseline.}
    \label{fig:Wiggling time domain} 
\end{figure}

\begin{figure}[htb!]
    \centering
    \includegraphics{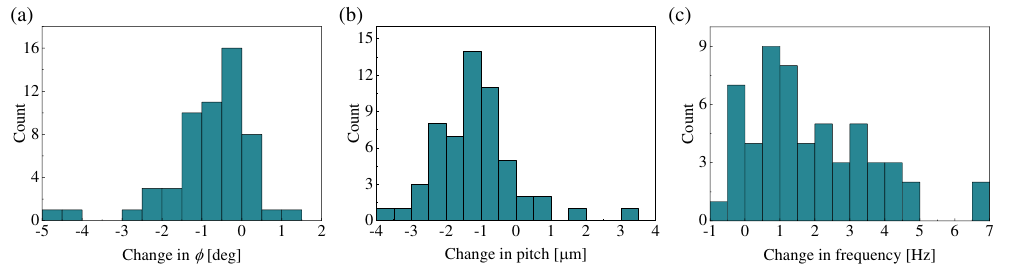}
    \caption{Changes in wiggling angle $\phi$, frequency, and pitch as the bacteria enter the nematic phase from the isotropic phase. We analyzed 56 trajectories out of the 63 shown in Fig.~\ref{fig:angle change}.}
    \label{fig:Wiggling change} 
\end{figure}

We first estimated baselines from the trajectories to find the wiggling angle $\phi$. Fig.~\ref{fig:Wiggling time domain}(a) shows a wiggling trajectory in the time domain. After identifying local extrema, we defined the baseline by interpolating midpoints between adjacent local maximum and minimum, as shown in Fig.~\ref{fig:Wiggling time domain}(b). Subtracting the baseline from $\theta(t)$ in Fig.~\ref{fig:Wiggling time domain}(a) gives $\delta \theta(t)$. We applied the same method for Fig. 4(a). We analyzed the $\delta \theta$ curves in spatiotemporal domains to determine the wiggling angle, pitch, and frequency.

Figure S7 shows how bacteria wiggling differs between the isotropic and nematic phases. 
Average values in each phase are $\phi_{\mathrm{I}} = 3.7 \pm 1.7$ deg, $\phi_{\mathrm{N}} = 2.9 \pm 1.0$ deg, $\mathrm{frequency}_{\mathrm{I}} = 4.6 \pm 1.5$ Hz, $\mathrm{frequency}_{\mathrm{N}} = 6.4 \pm 1.8$ Hz, $\mathrm{pitch}_{\mathrm{I}} = 5.5 \pm 1.1~\mu$m, and $\mathrm{pitch}_{\mathrm{N}} = 4.4 \pm 1.2~\mu$m.
A clear distinction with smaller wiggling angle but faster frequency is observed in swimming in the nematic phase with respect to the isotropic phase.

\newpage
\bibliography{References}